# LOAD BALANCING USING ANT COLONY IN CLOUD COMPUTING

Ranjan Kumar[1] and G Sahoo[2]

[1]Department of Computer Science & Engineering, C.I.T Tatisilwai, Ranchi, India
[2]Department of Information Technology, B.I.T Mesra, Ranchi, India

***ABSTRACT***

*Ants are very small insects.They are capable to find food even they are complete blind. The ants lives in their nest and their job is to search food while they get hungry. We are not interested in their living style, such as how they live, how they sleep. But we are interested in how they search for food, and how they find the shortest path. The technique for finding the shortest path are now applying in cloud computing. The Ant Colony approach towards Cloud Computing gives better performance.*

## 1. INTRODUCTION

Cloud Computing is very hot topic in IT field. Many researches are going on Cloud Computing. This is basically "on-demand" service. It means whenever we need for some applications or some software, we demand for it and we immediately get it. We have to pay only that we use. This is the main motto of cloud computing. Our desired application will present in our computer in few moment. Cloud Computing has basically two parts, the First part is of Client Side and the second part is of Server Side. The Client Side requests to the Servers and the Server responds to the Clients. The request from the client firstly goes to the Master Processor of the Server Side. The Master Processor are attached to many Slave Processors, the master processor sends that request to any one of the Slave Processor which have free space. All Processors are busy in their assigned job and non of the Processor get Idle. The process of assigning job from Master processor to the Slave processor and after completion the job, then returning from the Slave processor to the Master processor is just like Ant takes their food and return to their nest. The real ants left out pheromone while travelling. A pheromone is a chemical used for communication. Now we are moving from real ants to artificial ants. The artificial ants have some special characteristics which is not found in real ants, such as they are not completely blind, they have some memory called tabu. Now the artificial ants are used in cloud computing. The cloud computing is composed of three service models, five essential characteristics, and four deployment models.

The three service models are as follows.

Software as a Service (SaaA).
Platform as a Service (PaaS).
Infrastructure as a Service (IaaS).
The five essential charactersistics are as follows.

On-demand self service
Ubiquitous network access
Resource pooling
Rapid elasticity
Location independence
The four deployment models are as follows.
Private Cloud
Public Cloud
Community Cloud
Hybrid Cloud

Organization of this paper is as follows: Related work is discussed in section II. Proposed Ant Colony is discussed in section III. Experimental setup is discussed in section IV. Result is discussed in section V. And section VI gives conclusion.

## 2. RELATED WORK

Marco Dorigo and Luca Maria Gambardella [1] described about real and artificial ant. An artificial ant colony, that was capable of solving Travelling Salesman Problem. Real ants are capable of finding the shortest path from food source to the nest without using visual cues. Also, they are capable of adapting to changes in the environment, for example finding a new shortest path once the old one is no longer feasible due to a new obstacle. Zehua Zhang and Xuejie Zhang [2] described about Load balancing mechanism based on Ant Colony. They described about the function of Load balancing and how to distribute the workload in a cloud and to realize a high ratio of user satisfaction. They described the two characteristic of Complex Network and these two characteristics are considered for the move of the ants in the work, since the ants move more quickly towards that region where more resources found. They also described about Underload and Overload of load balancing methods. Sarayut Nonsiri and Siriporn Supratid [3] discussed about the ACO that allows fast near optimal solutions to be found. It is useful in industrial environments where computational resources and time are limited. Patomporn Premprayoon and Paramote Wardkein [4] discussed about the topological communication network design. They discussed about the backbone network and the Local Area Network (LAN), they give the formula of Total number of possible links in a single design. They discussed about the Reliability calculation using backtracking algorithm for correctly calculate the system reliability. They also discussed about the basic principle of ant colony and State Transition Rule in Ant Colony Optimization technique and Global updating rule. Zenon Chaczko, Venkatesh Mahadevan, Shahrzad Aslanzadeh and Christopher Mcdermid [7] discussed about the availability and load balancing in cloud computing. They discussed about the static and dynamic algorithms and the load balancing techniques to obtain measurable improvements in resource utilization and availability of cloud computing environment.

## 3. PROPOSED ANT COLONY

Marco Dorigo, first introduced the Ant System (AS) in his Ph.D thesis in 1992. Now it is one of the best optimization technique, which finds the shortest path. The deposition of pheromone and the ant move is approximately at the same speed and at the same rate. And that pheromone attracts another ants to move on same path. So, more ants move on same path have higher concentration of pheromone and the evaporation rate is very low on shorter path, that's why ants chooses the shorter path.

The probability with which ant $k$ currently at stage $i$ choosing to go to stage $j$.

$$p_{ij}^k(t) = \frac{[\tau_{ij}(t)]^\alpha [\eta_{ij}(t)]^\beta [A_p]}{\sum_{l \in J_i^k} [\tau_{ij}(t)]^\alpha [\eta_{ij}(t)]^\beta}$$

Where,

$\tau_{ij}$ = Pheromone trail

$\eta_{ij}$ = Heuristic value

$\alpha$ = Parameter which determines the relative influence of the pheromone trail.

$\beta$ = Parameter which determines the relative influence of the pheromone trail.

$A_p$ = Amount of pheromone

The proposed Algorithm is defined as follow.

Step 1 : Randomly select a Job Schedular.

Step 2 : Job Schedular Schedules job to different web services.
   While Job is not schedule to web services
      Repeat steps 3 & 4.

Step 3: Job checks its surrounding area for availability of web services with Probability,

$$p_{ij}^k(t) = \frac{[\tau_{ij}(t)]^\alpha [\eta_{ij}(t)]^\beta [A_p]}{\sum_{l \in J_i^k} [\tau_{ij}(t)]^\alpha [\eta_{ij}(t)]^\beta}$$

Step 4 : if

   Web server is available

   Then
      Acquire web server
   Else
      Go to step 3

Step 5 : Return to the Job Schedular.

Step 6 : After completition kill the job.

Step 7 : End

The Web Services have some amount of load at any time, since non of the processor get idle. The decision point makes ants to realize the Load of different Web Services.

## 4. EXPERIMENTAL SETUP

To evaluate the performance of Ant Colony, the results were simulated in Window 7 basic (64-bit), i3 processor, 370 M processor, 2.40 GHz of speed with memory of 3 GB and language used C++. There are 10 job sechedulers and 44 different web services. The job secheduler sechedules the different jobs to the different web services. The number of ants in this simulation varies from 1 to 1000. These ants deposit some amount of pheromone in there move.

## 5. RESULT

We have experimented by taking different amount of number of ants. The amount of pheromone varies between 0 to 1. The table I shows the number of ants and the amount of pheromone deposited.

Table I

| No. of Ants | Amount of Pheromone |
| --- | --- |
| Upto 10 | 0.01-0.10 |
| Upto 20 | 0.10-0.15 |
| Upto 30 | 0.15-0.17 |
| Upto 50 | 0.17-0.19 |
| Upto 90 | 0.20-0.30 |
| Upto 100 | 0.35-0.45 |
| Upto 200 | 0.50-0.65 |
| Upto 300 | 0.65-0.75 |
| Upto 600 | 0.75-0.85 |
| Upto 1000 | 0.85-1.00 |

From the table I, we see that as the number of ants increases, the amount of pheromone also increases, Since most of the ants uses the same path. The figure I shows the graph of Table I.

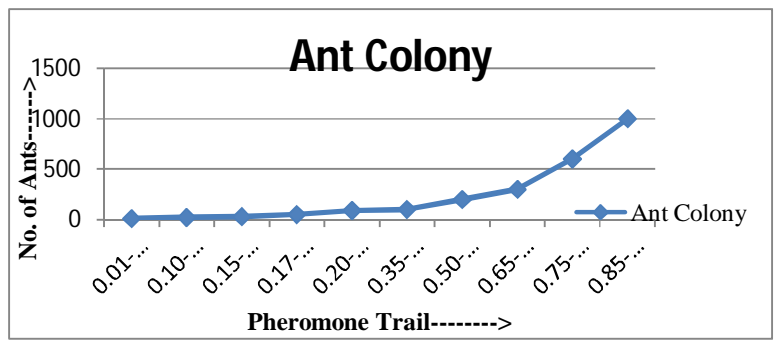

Figure I. Ant Colony in respect of Ants & Pheromone Trail

## 3. CONCLUSIONS

In this paper, we have proposed a method for load balancing. In which we emphasis on deposition of pheromone. Here we see that when a node with minimum load is attracted by most of the ants gives result to the maximum deposition of pheromone.

**Authors**

Ranjan Kumar  received  M.Tech degree in Computer Science from B.I.T Mesra, Ranchi. He has one year teaching experience. His research interests include cloud computing, Algorithm and compiler.

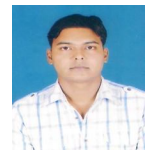